\documentclass[manuscript]{aastex}
\usepackage{natbib}
\usepackage{booktabs}
\usepackage{threeparttable}
\begin{document}

\title{Time-dependent occurrence rate of electromagnetic cyclotron waves in the solar wind: evidence for effect of alpha particles?}

\author{G. Q. Zhao\altaffilmark{1,2}, H. Q. Feng\altaffilmark{1}, D. J. Wu\altaffilmark{3}, Y. H. Chu\altaffilmark{4}, and J. Huang\altaffilmark{2}}
\affil{$^1$Institute of Space Physics, Luoyang Normal University, Luoyang, China}
\affil{$^2$CAS Key Laboratory of Solar Activity, National Astronomical Observatories, Beijing, China}
\affil{$^3$Purple Mountain Observatory, CAS, Nanjing, China}
\affil{$^4$Institute of Space Science, National Central University, Chungli, Taiwan}

\begin{abstract}
Previous studies revealed that electromagnetic cyclotron waves (ECWs) near the proton cyclotron frequency exist widely in the solar wind, and the majority of ECWs are left-handed (LH) polarized waves. Using the magnetic field data from the STEREO mission, this Letter carries out a survey of ECWs over a long period of 7 years, and calculates the occurrence rates of ECWs with different polarization senses. Results show that the occurrence rate is nearly a constant for the ECWs with right-handed polarization, but it varies significantly for the ECWs with LH polarization. Further investigation of plasma conditions reveals that the LH ECWs take place preferentially in a plasma characterized by higher temperature, lower density, and larger velocity. Some considerable correlations between the occurrence rate of LH ECWs and the properties of ambient plasmas are discussed. The present research may provide evidence for effect of alpha particles on generation of ECWs.
\end{abstract}


\section{Introduction}

In recent years, a series of studies showed that coherent low frequency electromagnetic cyclotron waves (ECWs) with a typical frequency of 0.1$-$0.5 Hz at 1 AU can be detected widely in the solar wind \citep{jia09p05,jia10p15,jia14p23,boa15p10,gar16p30,wic16p06}. These waves are characterized by narrow band and have frequencies near the proton cyclotron frequency. They are transverse waves and propagate mainly in the directions quasi-parallel (or antiparallel) to the ambient magnetic field. They can be sporadic in occurrence with a median duration of 51.5 s \citep{jia09p05}, or appear in clusters with durations exceeding 10 min \citep{jia14p23}. Their polarization senses can be left-handed (LH) or right-handed (RH) with respect to the magnetic field in the spacecraft frame, and more ECWs were found to be LH polarized waves with percentage of 64\% \citep{jia09p05} or 55\% \citep{jia14p23} at 1 AU.

Theoretically, many mechanisms can contribute to the generation of the ECWs. They are related to plasma instabilities driven by temperature anisotropies and/or differential velocities between ion populations \citep{gar93,lix00p83,luq06p01,ver13p63,omi14p42}. A plasma with perpendicular temperature ($T_{\perp}$) larger than parallel temperature ($T_\parallel$) may amplify ion cyclotron waves that are inherently LH polarized, while a plasma with converse temperature anisotropy ($T_{\perp} < T_\parallel$) can excite magnetosonic waves that are RH polarized in the plasma frame. When ion beam/core relative flow speed is large sufficiently (typically exceeding the Alfv\'en speed; \citet[][p166]{gar93}), the plasma may generate ion cyclotron waves or magnetosonic waves depending on beam parameters. For details of the mechanism associated with differential flow of protons, one can also refer to the literature \citep[e.g.,][]{abr79p53,dau98p13,dau99p57,gol00p53}. The combined effects of temperature anisotropies and proton differential flows have also been discussed in the recent year, in the case of slow solar wind \citep{gar16p30} as well as descending part or trailing edge of fast solar wind \citep{wic16p06,jia16p07}. Their results tend to suggest that the main driver of instabilities is temperature anisotropies, but proton differential flows provide additional free energy and amplify the wave growth in the solar wind \citep{wic16p06,jia16p07}.

In this Letter, we report our finding that the LH ECWs and RH ECWs have significantly different behaviors in their time-dependent occurrence rates as well as preferential plasma conditions, which may provide indication or important constraint on the mechanisms of generating ECWs in the solar wind. The data and analysis methods used in this Letter are described in Section 2. Section 3 presents the results, and Section 4 gives our discussion and summary.

\section{Data and analysis methods}
The data used in the present Letter are from the STEREO-A spacecraft, which has orbits near 1 AU in the ecliptic
plane and can provide continuous magnetic field data with resolution of 8 Hz as well as plasma data with resolution of 1 min \citep{kai08p05,luh08p17,gal08p37}.  Based on the magnetic field, we perform a survey of ECWs and calculate their occurrence rates over the period between 2007 and 2013. An automatic wave detection procedure is employed to identify ECWs. The procedure is developed by \citet{zha17p79} and mainly consists of three steps for magnetic field data in some time interval of 100 s. The first step is to obtain the normalized reduced magnetic helicity that takes values in the range from $-1$ to 1 \citep{mat82p11,gar92p03,hej11p85}. The magnetic helicity is actually a spectrum with resolution of 0.01 Hz, and the spectrum values are examined in the frequency range from 0.05 to 1 Hz. If the spectrum has positive values $\geq$ 0.7 or negative values $\leq -$0.7 in some frequency band with minimum bandwidth of 0.05 Hz, the second step will be carried out to identify enhanced power spectrum requiring transverse wave power three times larger than the background power in the same frequency band; the background power is obtained via fitting the entire transverse power spectrum with a power law. A wave amplitude criterion of 0.1 nT is also set, which completes the third step. During the process a Hamming window and a band-pass filter are used to reduce edge effects and determine a wave amplitude \citep{bor07p04,wil09p06}. The procedure can give the time intervals of ECW occurrence, and their time occurrence rates accordingly. It can also directly give the polarization senses of the waves based on the sign of spectrum values of magnetic helicity.

Note that the polarization is described in the spacecraft frame throughout the paper except that we point out the plasma frame. The polarization will reverse in the two different reference frames for ECWs propagating toward the Sun \citep[e.g.][]{jia09p05,gar16p30}. This is due to the presence of large Doppler shift resulting from fast movement of the solar wind relative to the (approximately standing) spacecraft. The speed of the solar wind is typically 5$-$8 times of the Alfv\'en speed, which is also much greater than the phase velocity of the ECWs, i.e., typically the Alfv\'en speed \citep{jia09p05,jia10p15}. The Doppler shift can be estimated via the relation $f_{sc} = f_{sw}(1+\frac{V_{sw}}{V_A}\hat{\textbf{k}}\cdot\hat{\textbf{V}}_{sw})$ introduced by \citet{jia09p05}, where $f_{sc}$ and $f_{sw}$ are the wave frequency in the spacecraft frame and in the plasma frame, respectively, $\textbf{k}$ denotes the wave propagation vector, $\textbf{V}_{sw}$ is the solar wind velocity, and $V_A$ is the Alfv\'en speed. It is clear that the second term dominates with consideration of $\hat{\textbf{k}}\cdot\hat{\textbf{V}}_{sw} \simeq 1$ (for ECWs propagating away from the Sun) or $\hat{\textbf{k}}\cdot\hat{\textbf{V}}_{sw} \simeq -1$ (for ECWs propagating toward the Sun).

\section{Results}
Figure 1 presents the most important finding of this Letter, in which the occurrence rates for each month are plotted. In the figure the red line shows the occurrence rates of ECWs with LH polarization and blue line displays those with RH polarization. One may first find that the occurrence rate of LH ECWs is larger than that of RH ECWs for most of months, which is compatible with previous result that more ECWs are LH polarized waves in the solar wind \citep{jia09p05,jia10p15,jia14p23,boa15p10}. In particular, the occurrence rate of LH ECWs fluctuates considerably in a wide range from about 0.5\% to 2\%, it exceeds 2.5\% in some months. The occurrence rate of RH ECWs, however, just shows weak fluctuation around 0.46\% (mean value, with a standard deviation of 0.22\%), and may be seen approximately as a constant relative to the occurrence rate of LH ECWs. In addition, it should be noted that minimum of the occurrence rate of LH ECWs is comparable to that of RH ECWs.

In order to understand the implication of the result presented in Figure 1, we investigate the local plasma characteristics associated with occurrence of ECWs, as well as dependence of their occurrence rates on the ambient plasma properties. Results reveal the preferential plasma conditions favoring the LH ECWs and considerable dependence for the LH ECWs. Figure 2 displays plasma parameters with respect to months, where panels from top to bottom correspond to the proton temperature ($T_p$), proton density ($N_p$), and proton velocity ($V_p$). (The plasma data in the first month of 2007 are not available.) In each panel the red line is for median of a plasma parameter associated with LH ECWs, while the black line is for median of the plasma parameter for all plasmas (referred to as ``ambient median" for convenience). The plasma parameter associated with ECWs refers to the plasma data almost simultaneously arising with the ECWs; here an averaging operation is made for the plasma data over each time interval of 100 s once an ECW is found in the same interval. On the contrary, the plasma parameter for all plasmas refers to all the plasma data irrespective of whether the ECWs are in presence or not. One can find that the medians associated with the waves vary with trends similar to those of the ambient medians. In particular, the median of proton temperature (density) as well as velocity is predominantly larger (smaller) than the corresponding ambient median. This result should be interesting and tends to imply that the high temperature, low density, and large velocity are preferential plasma conditions for generation or survival of LH ECWs in the solar wind.

Furthermore, it seems to be fulfilled that a plasma with higher temperature, lower density, and larger velocity will favor a higher occurrence rate of LH ECWs, since these quantities show positive or negative correlations by comparing the time series between Figures 1 and 2. The correlations are particularly high for proton temperature and velocity.
Figure 3 presents scatter plots of the occurrence rate of LH ECWs against ambient medians of proton temperature (left panel) and velocity (right panel). The line with positive slope in each panel of Figure 3 is the best linear fit to the scatter data. As shown, the correlations are considerable, with their correlation coefficients ($C$) as high as close to 0.8.

As for the case of RH ECWs (not shown), the above preferential plasma conditions for the wave generation as well as dependence of occurrence rate of the waves on ambient plasma properties are not clear; the medians of the plasma parameters for the RH ECWs vary around the ambient medians, and the correlation between the wave occurrence rate and ambient temperature as well as velocity is also negligibly small, with $C < 0.2$.

\section{Discussion and summary}
ECWs are common wave activities in the solar wind \citep{jia09p05,jia14p23}. Using the data from the STEREO-A spacecraft and the method developed by \citet{zha17p79}, this Letter first carries out a survey of the occurrence rates of ECWs in the solar wind for each mouth between 2007$-$2013. Results show that the occurrence rate of LH ECWs is larger than that of RH ECWs for most of months. Moreover, the occurrence rate of LH ECWs fluctuates considerably in a wide range, while the occurrence rate of RH ECWs tends to be a constant approximately. In addition, the minima of occurrence rates are comparable for LH and RH ECWs. Preferential plasma conditions favoring LH ECWs and considerable dependence of the occurrence rate of LH ECWs on the ambient plasma properties are revealed, which may provide indication on the mechanisms of generating ECWs in the solar wind.

On the basis of the results in Figures 2 and 3, one may speculate that high-speed solar wind streams are relevant to answer the question what factor results in the difference of occurrence rates between LH and RH ECWs presented in Figure 1. A lot of studies have shown that the plasmas in high-speed streams are characterized by higher temperature, lower density, and larger velocity than those in slow-speed flows \citep{bur74p17,gos78p01,cra02p29}, which first provides preferential conditions contributing to a higher occurrence rate of LH ECWs according to the present research. Moreover, minor ions such as alpha particles in high-speed streams, flow generally faster than protons, therefore forming differential flow with velocity on the order of the Alfv\'en speed \citep{mar82p35,mar91p52,kas08p03}. We believe that the presence of such differential flow could offer a specific mechanism on generating LH ECWs in the solar wind. This idea is also supported by observation of alpha particles. Figure 4 (left panel) is scatter plot of the occurrence rate of LH ECWs versus ambient median of alpha$-$proton drift velocity ($V_d$) in each month. The alpha data are available intermittently from February in 2007 to December in 2010, and have a low resolution of 10 minute for the STEREO at present. Nevertheless, a positive correlation with coefficient exceeding 0.7 is shown. For the sake of comparison, the result for RH ECWs is also plotted in Figure 4 (right panel). The correlation for RH ECWs is small, with $C = 0.19$.

Theoretically, the significance of effect of alpha$-$proton differential flow on proton temperature anisotropy instabilities has been demonstrated via hybrid simulations and linear Vlasov$-$Maxwell theory \citep{hel06p07,pod11p41}. In particular, the results from \citet{pod11p41} show that the alpha$-$proton differential flow causes instability with $T_{\perp} > T_\parallel$ to preferentially generate ion cyclotron waves propagating away from the Sun, and it causes instability with $T_{\perp} < T_\parallel$ to preferentially generate magnetosonic waves propagating toward the Sun. Note that although magnetosonic waves are RH waves in the plasma frame, these waves shall also appear as LH waves in the spacecraft frame due to the large Doppler shift as described in Section 2. In one word, proton temperature anisotropies generate the observed LH ECWs when one takes into account the differential flow of alpha particles relative to the protons \citep{pod11p41}.

The present discussion tends to imply local sources of ECWs driven mainly by proton temperature anisotropies that are common in the solar wind \citep{mar82p52,mar04p02,mat12p73}. The temperature anisotropies can generate ion cyclotron waves or magnetosonic waves via instabilities. These waves may propagate toward or away from the Sun with a comparable probability. The second version may be that almost all of the waves propagate away from the Sun, where the waves are composed of ion cyclotron waves and equal amount of magnetosonic waves in principle \citep{gar16p30,jia16p07}. Both cases above lead to a similar occurrence rate for LH and RH ECWs, which may be the reason why LH and RH ECWs have their comparable minima of occurrence rates shown in Figure 1. However, the presence of high-speed streams and therefore differential flow of alpha particles would change the situation above; the differential flows have an important effect on the proton temperature anisotropy instabilities and causes the instabilities preferentially generating LH ECWs \citep{pod11p41}. In this regard, it becomes easy to understand why occurrence rates for LH and RH ECWs are different and more ECWs are LH waves in the solar wind.  

In summary, this Letter finds significant differences in behavior of occurrence rates between the LH ECWs and RH ECWs. The occurrence rate for each month is nearly a constant for the RH ECWs over the period of 7 years, but it varies significantly for the LH ECWs over the same period. The plasma with a higher temperature, lower density, and larger velocity favors the LH ECWs, but there seems to be no preferential conditions for the RH ECWs. Further analysis indicates that the present finding is consistent well with the theory for effect of differential flow of alpha particles on generation of ECWs. This finding hence probably is an evidence for the effect concerning alpha particles. Further studies related to instability simulations are needed, and the parameters found here could constrain initial conditions for the simulations to confirm the speculations in the present Letter.

\acknowledgments
This research was supported by NSFC under grant Nos. 41504131, 41674170, 41531071, 11373070, and the Key Laboratory of Solar
Activity at CAS NAO (KLSA201703). This research was also sponsored partly by the Plan For Scientific Innovation Talent of Henan Province. The authors thank NASA/GSFC for the use of data from STEREO, which are available freely via the Coordinated Data Analysis Web (http://cdaweb.gsfc.nasa.gov/cdaweb/istp$_-$public/).


\begin{figure}
\epsscale{0.9} \plotone{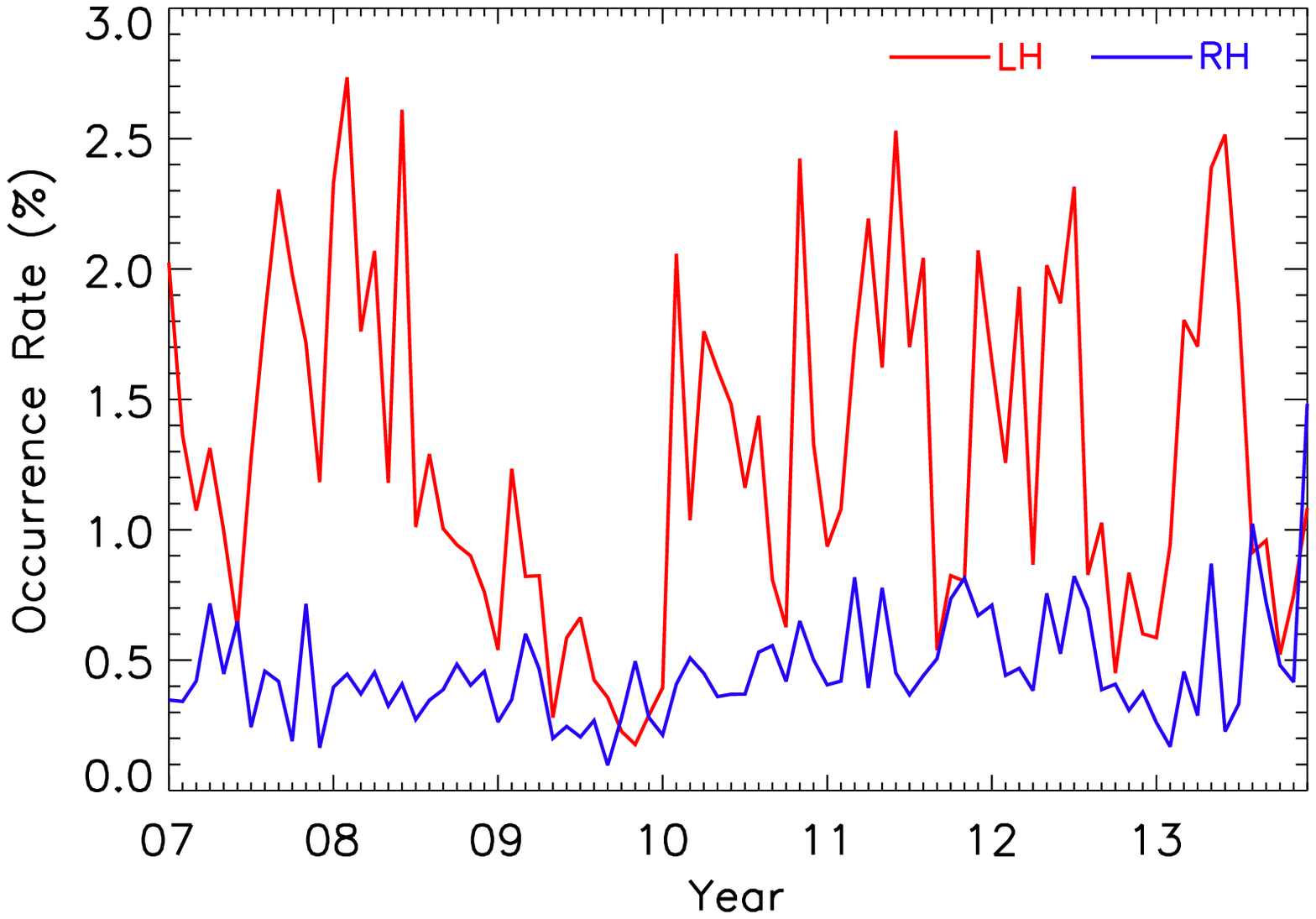} \caption{Time occurrence rates with respect to months in the years between 2007 and 2013. It is clear that the occurrence rate for LH ECWs (red line) varies significantly, while it is nearly a constant for RH ECWs (blue line).}
\end{figure}

\begin{figure}
\epsscale{0.9} \plotone{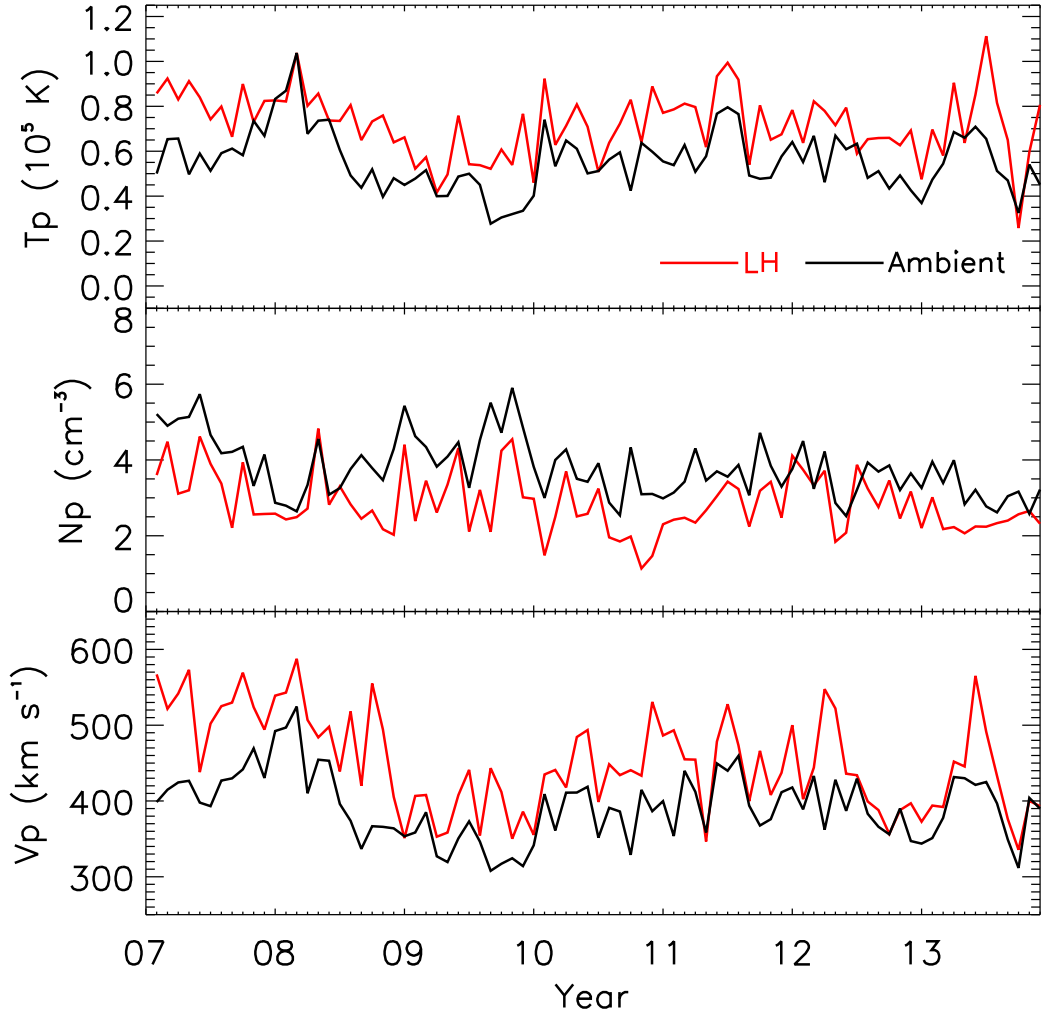} \caption{Medians of proton temperature ($T_p$), proton density ($N_p$), and proton velocity ($V_p$) with respect to months in the years between 2007 and 2013. The red line is for median associated with LH ECWs, while the black line is for median with all plasmas in each panel.}
\end{figure}

\begin{figure}
\epsscale{0.9} \plotone{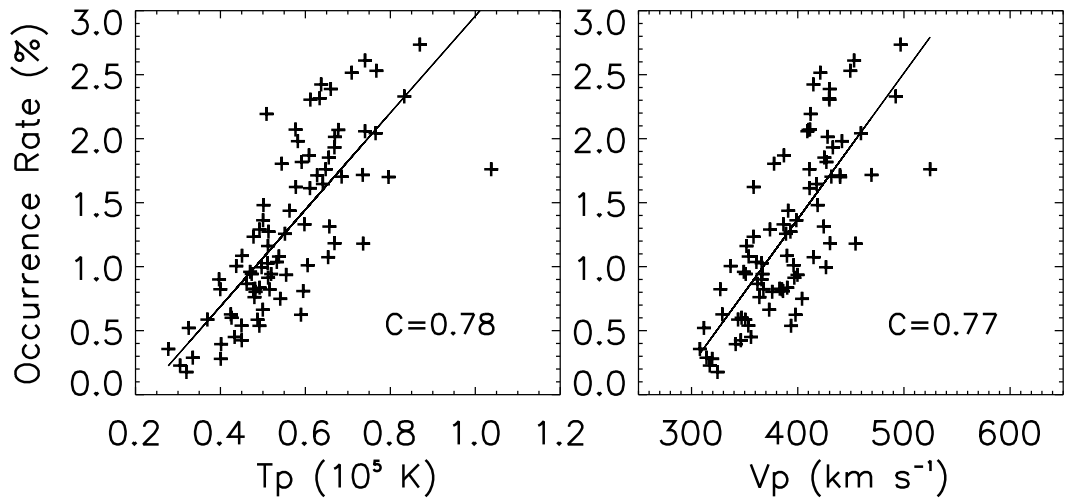} \caption{Scatter plots of occurrence rate of LH ECWs against ambient medians of proton temperature (left panel) and velocity (right panel). The line in each panel denotes the best linear fit, with the correlation coefficient (C) in the lower right corner of the panel.}
\end{figure}

\begin{figure}
\epsscale{0.9} \plotone{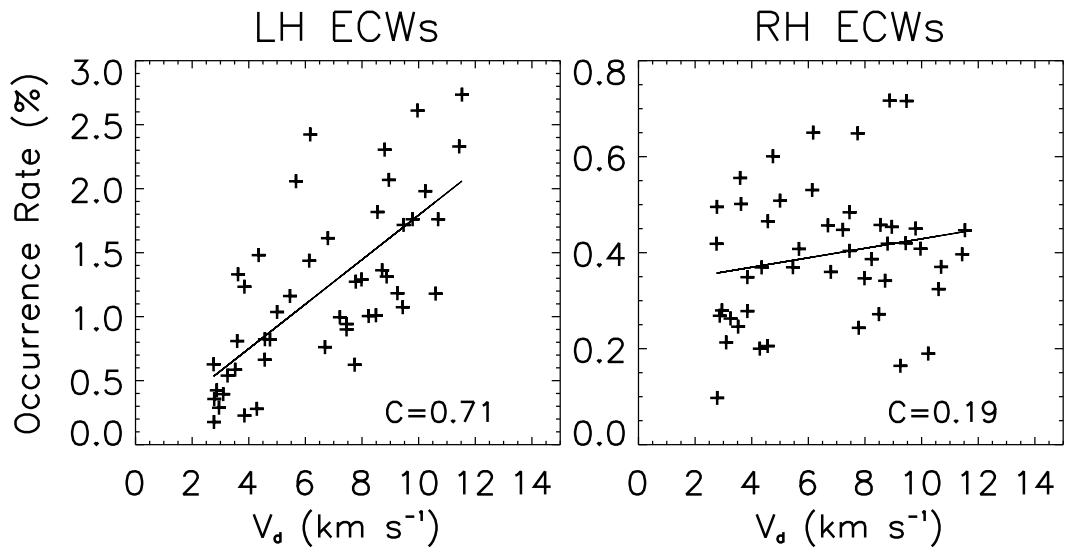} \caption{Scatter plots of occurrence rates of LH ECWs (left panel) and RH ECWs (right panel) against ambient median of alpha$-$proton drift velocity ($V_d$). The line in each panel denotes the best linear fit, with the correlation coefficient (C) in the lower right corner of the panel.}
\end{figure}

\end{document}